# Evolution of Fermi Level State Density in Ultrathin Films Near the Two–Dimensional Limit: Experiment and Theory

R. Batabyal, A. H. M. Abdul Wasey, J. C. Mahato, Debolina Das, G. P. Das and B. N. Dev*

*Department of Materials Science, Indian Association for the Cultivation of Science,*

*2A and 2B Raja S. C. Mullick Road, Jadavpur, Kolkata 700032, India*

## Abstract

Electronic density of states (DOS) at Fermi level has been investigated in ultrathin Ag films grown on Si(111)-(7×7) down to the two–dimensional limit of a single atomic layer. Measurement of DOS at Fermi level by scanning tunneling spectroscopy shows an approximate $(1-\gamma/d)$ dependence, where $\gamma$ is a constant and $d$ is the film thickness. The results are explained in the light of an analytical theory as well as our density functional theory (DFT) calculations. DFT results also show that in the proximity of the interface the DOS values of the film and the substrate are mutually affected by each other.

Properties of lower dimensional structures of a material are known to differ from the corresponding bulk or three-dimensional (3-D) counterpart. Lower dimensional structures have led to important discoveries and advanced technological applications. Recent research activities additionally show that in the dimensional crossover regimes, when a structure begins to evolve from a given lower dimension towards a higher dimension, many properties also evolve with the system size. With a scanning tunneling microscope, structures in the 0-D to 1-D crossover regime were fabricated by adding Au atoms in a linear chain of different lengths and their size-dependent electronic structures were determined [1]. In the 2-D to 3-D crossover regime near the 2-D limit, i.e., for one- to a-few- atomic layer thick films, changing superconducting properties and electronic transport properties have been observed as the system size increased from 2-D towards 3-D. Superconductivity in ultrathin Pb films on Si(111) near the 2-D limit (two-to-five atomic layer thickness) has shown interesting variation in superconducting gap and superconducting transition temperature [2]. In one-to-five atom thick Ag layers on Si(111), a negative differential resistance (NDR) behavior has been observed in electron tunneling transport. The sample bias voltage at which NDR appears has been found to depend on the number of atomic layers in the structure [3]. Ultrathin magnetic films over a few atomic layer thicknesses have shown a rapid decrease in Curie temperature with decreasing film thickness [4]. Electronic density of states at Fermi level has shown size-dependent change for structures in the 0-D to 2-D crossover regime [5]. On the theoretical side, a simple free electron model has shown how electronic properties evolve in various dimensional crossover regimes [5] indicating tunability of various properties in appropriately fabricated structures. This elucidates the importance of both experimental and theoretical investigations of structures in various dimensional crossover regimes.

For solid materials, a very important parameter is the electronic density of states (DOS) at Fermi energy ($E_F$), which determines a large number of properties of solids. Electronic DOS at $E_F$ in ultrathin films is expected to show a thickness dependence [5 - 7]. Properties like electronic specific heat [6], spin susceptibility [6], Thomas-Fermi screening length [8], superconducting transition temperature [9 - 11], magnetic anisotropy [12] and many other properties, which depend on the value of electronic DOS at $E_F$, would vary with film thickness. This means that all these properties can be tailored by appropriate electronic-structure engineering. The present work deals with the thickness dependence of electronic structure in ultrathin films.

In this Letter, we report on the determination of electronic DOS at $E_F$ in ultrathin Ag films on Si(111) in the 2-D to 3-D cross over regime near the 2-D limit. Ag on Si is an ideal system used for various investigations in surface physics. We have determined electronic DOS at $E_F$ for 1-6 atomic layers of Ag on Si(111) by scanning tunneling spectroscopy (STS) and compared the results with our density functional theory (DFT) calculations. The results are in good agreement with the DFT calculations as well as with an analytical theory [7] and a free electron model simulation [5].

Ag thin films have been grown on Si(111)-7×7 surfaces under ultrahigh vacuum (UHV) condition and investigated using *in-situ* scanning tunneling microscopy (STM) and STS in an equipment similar to that in Ref. [13]. Ag was deposited on atomically clean Si(111)-7×7 surfaces at room temperature (RT) at a coverage of 2 monolayers (MLs). (Here 1 ML is equivalent to $1.5 \times 10^{15}$ atoms/cm$^2$ as in refs. [14] and [15]). This has produced flat-top islands of different heights. We have chosen a large area island of six atomic-layer (6-AL) height. Applying a voltage pulse (+7V was applied for 1s keeping the feed-back loop on) from a STM (VTSTM, Omicron nanotechnology) tip we have created a "V"-shaped groove on this island. On the slope of this groove we have accessed different heights and made STS measurements at different points corresponding to different film thicknesses. (Earlier, wedge-shaped samples were used to obtain results for different thicknesses on a single sample [16, 17]). This strategy was applied for the following reasons.

Deposition of 1 ML Ag on Si(111) at low temperature followed by annealing at RT [14] or directly at RT [15] produces two-atom tall Ag islands on a wetting layer of a single atomic layer of Ag. Deposition at higher coverages, 2 ML onwards, produce percolated structures of Ag(111) islands with preferential heights containing two-, four-, six- ... atomic layers of Ag on a wetting layer, although islands of other heights are also present [15,18]. Such a system is suitable for the investigation of electronic DOS at the Fermi level by STS for a range of film thicknesses in a single preparation. However, while making a measurement of DOS on a flat-top island, one needs to ensure that the lateral dimension of the island would suffice for the island to be considered two-dimensional [5]. (Experiments in refs. [2] and [3] were carried out on islands of different thicknesses). In order to avoid any lateral size dependence in the measurement of DOS on islands of different heights, we have made measurements on a single island by creating a "V"-shaped groove and accessing different film thicknesses on the slope of the groove. We have chosen a single island for measurements due to another reason. That is, the electronic structure of individual islands of different heights may be affected by different structural relaxations and quantum confinement [18]. Measurement on a single island can avoid such effects. A third reason for this strategy is that the structure of a one-atomic layer (1-AL) Ag film on Si(111), i.e. the wetting layer, is not well defined and not like Ag(111); a 2-AL film has not been found to form in the growth process followed here [14 - 15, 18]. However, under an island Ag is (111) down to the first Ag layer at the interface [15, 19 - 21].

Deposition of 2 ML Ag provides connected islands of higher thicknesses as in ref. ref. [3] and [15]. A part of the Ag film is shown in Fig. 1(a). The height scan of the film along the marked line is shown in Fig. 1(b). The thickest part of the film contains 6 atomic layers (6-AL) of Ag (1-AL + 5-AL). As we mentioned before, we have created a groove on a particular island (see Fig. 2(a)) using a STM tip voltage pulse on the top of the island. *I-V* measurements were carried out at different points from the center of the groove towards the island top, thus accessing different thicknesses of the Ag film. Fig. 2(b) shows the actual positions where the measurements were made. The positions of *I-V* measurements are schematically shown in Fig. 2(c). The *I-V* measurements from position 1 to 6 imply that the spectra are taken for different thicknesses of the Ag layer. Fig. 2(d) shows the actual depth profile and its derivative across the groove. Fig. 3(a) shows the *I-V* curves obtained from positions 1 to 6. Fig. 3(b) shows the corresponding (*dI/dV*)/(*I/V*) which is proportional to the local electronic DOS (LDOS) curves. DOS at Fermi level are the values of LDOS at *V=0*. These values from Fig. 3(b) are plotted in Fig. 4 vs. film thickness, which was obtained by STS measurement on the slope of the groove. Each point in Fig. 4 is an average of five sets of data. The data for 2-AL case is missing in the experimental results. The points on the slope of the "V" groove were selected manually (Fig. 2(b)). The heights of these points were measured afterwards. As the slope near the bottom of the groove is very steep (Fig. 2(b)), the 2-AL point was missed in manual selection. We try to explain this result based on the existing analytical theory and our numerical calculation based on free-electron theory as well as *ab-initio* density functional theory (DFT).

For a thin metal slab of infinite lateral dimension, the density of states at Fermi energy, $\rho(E_F)$ should depend on the slab thickness [7]. In a free-electron model with finite potential depth in the direction perpendicular to the surfaces of the film, $\rho(E_F)$ can be written as [7],

$$\rho(E_F) = \rho^0(E_F^0)\left(1 - \frac{\gamma^0}{d_z}\right) \qquad (1)$$

where, $\rho^0(E_F^0)$ is the bulk value of DOS at Fermi energy, $d_z$ is thickness of the infinitely large metal slab, and $\gamma_o$ is a constant. The details are in the supplementary material [22]. Now, we try to fit our experimental data for LDOS ($E_F$) normalized to its bulk value (or saturated bulk value) with thickness ($d_z$) of the film. Experimental data and the theoretical fitted curve to Eqn. (1) (with $\gamma^0$=0.2077 nm) are shown in Fig. 4. Therefore, analytical solution from the free-electron model has a good agreement with our experimental result. Our numerical calculations from the free electron model, on the evolution of DOS from a 2-D to a 3-D system [5], show a trend consistent with the experimental results. The DFT results are discussed below.

The experiment is on thin films grown on a substrate, as in the most cases of experiments on ultrathin films while the theoretical trends shown above are for free-standing films. We have carried out more sophisticated DFT calculations [23, 24] in order to understand the thickness dependence of the DOS at Fermi energy of ultrathin Ag films on Si (111). We found good agreement with experiments for film thicknesses 3-AL onwards. The discrepancies for the thinnest films will be explained latter. We first construct the model supercell of Ag(111) layers on a Si(111) substrate. Although there is a ~ 25% lattice mismatch between Ag and Si, epitaxial growth of Ag on Si is possible via coincidence site lattice matching as *4$a_{Ag}$ ≈ 3$a_{Si}$* ($a_{Ag}$ and $a_{Si}$ are lattice constants of Ag and Si respectively). The coincidence site lattice matching between 4×4 surface unit cells of Ag and 3×3 surface unit cells of Si reduces the lattice mismatch to ~ 0.3% [14]. That is how epitaxial Ag(111) islands are formed in Ag deposition on Si(111)-(7×7) at RT [15]. Cross sectional transmission electron microscopy for Ag deposition at RT has also shown abrupt Ag/Si interface with Ag adopting its natural lattice constant from the very interface showing *4$a_{Ag}$ = 3$a_{Si}$* relationship[19]. Thus at the interface every fifth atom of Ag would sit exactly above the fourth atom of Si. We have designed a supercell of Ag/Si(111) system for DFT study by placing Ag atoms onto the Si(111) substrate as shown in Fig. 5(a). The Si(111) slab constituting four bilayers of Si(111) represents the substrate. A Ag(111) layer is placed on one face of Si(111) while the other side of the Si(111) slab has been passivated with Hydrogen atoms to avoid the dangling bond states of Si(111) surface atoms. Furthermore, to study electronic structures for different thicknesses of Ag on Si(111)we have placed more number of Ag(111) atomic layers in fcc stacking on Si(111). More details of DFT calculations are given in the supplementary material [22]. Fig. 5(b) shows the computed DOS and Fig. 5(c) shows the DOS around the Fermi level (*V=0*). In Fig. 4 we notice that except for 1-AL and 2-AL thickness the DFT results agree well with experiment. (This kind of strong deviation of properties for the thinnest films was also observed in the investigation of superconductivity in Pb thin films on Si substrate. For the thinnest (2-AL) film, the influence of the Si substrate was found to affect the superconducting gap and transition temperature [2]. 1-AL film is not formed in this case [2]). We shall later discuss the possible reasons of the deviation of DOS at Fermi energy obtained from the DFT calculation for the case of 1-AL and 2-AL Ag layer on Si in the next paragraph. It should be mentioned here that in experiments, 1-AL film (i.e., the wetting layer) has not been found to be Ag(111) and 2-AL films are not formed [14, 15, 18]. However, under the islands all layers are Ag(111) [15, 19, 20, 21].

Layer-projected DOS of 1-AL, 2-AL, 3-AL and 4-AL Ag on four bilayers of Si(111)-(1×1) are shown in Fig. S1 [22]. Our DFT calculations show that near the interface, both Ag and Si electronic structures are modified. In fact Si does not show a band gap until the fourth atomic bilayer from the interface. Metal induced gap states (MIGS) appear into the band gap of Si [25]. Electronic states of the first Ag layer at the Ag/Si interface are also strongly modified. To find an answer to the question why the DOS at Fermi energy of 1-AL Ag obtained from DFT calculations, deviate so much from that obtained from the experiment, we analyze our DFT results in details.

The experimental situation is a bit different from the way DFT calculations have been performed. The experiments were performed on the wedge of a 'V'- shaped groove created on top of a flat-top 6-AL Ag island via applying tip pulse-voltage. Therefore, the first experimental point (as shown in Fig. 4) which represents the DOS at Fermi energy for first Ag layer, is located at the bottom of a 6-AL Ag island. This layer does not represent the 1-AL Ag on Si(111) as constructed for DFT calculation. The results obtained from the DFT calculation for the interface Ag layer, i.e. the first Ag layer of a 6-AL island, would be appropriate to compare with the experimental results. Due to computational limitations, we are not able to perform DFT calculations for 6-AL Ag island on Si. However, DFT calculations have been performed up to 5-AL Ag island on Si. Fig. S2(b) [22] shows that the DOS at Fermi energy of the interface Ag layer oscillates to saturate to the value ~ 2.3 as the thickness increases. Therefore, the DOS at Fermi energy for 1-AL Ag may be taken as ~ 2.3, instead of the calculated value of ~ 3.4 for 1-AL Ag, for comparison with the experimental result. When the calculated value is scaled down by the factor (2.3/3.4 or 0.676), the first point obtained from the DFT calculation (as shown in Fig. 4) would have a smaller value. This reduced value of DOS at Fermi energy for 1-AL Ag layer is shown as a star with a circle in its center in Fig. 4. This point is closer to the experimental value. As the experimental data point for 2-AL is not available, the DFT result cannot be compared here. In any case, in earlier experiments for Pb on Si, up to 2-AL thickness, the results have been found to be influenced by the proximity of the Si substrate [3].

In conclusion, we have demonstrated the dependence of electronic DOS at Fermi energy on layer thickness of an ultrathin Ag film on Si(111). In a large area island of 6-AL thickness, we have determined this dependence. Applying a voltage pulse from a STM tip on this island we have created a "V"-shaped groove. On the slope of this groove we have accessed different heights and carried out STS measurements at different points corresponding to different film thicknesses. In order to avoid any lateral size dependence in the measurement of DOS on islands of different heights, we have made measurements on a single island by accessing different film thickness on the slope of the groove made on it. Our results show that DOS at Fermi energy depends on the film thickness. DOS at Fermi energy increases and tends to saturate to its bulk value as the film thickness increases. Our experimental results are explained by density functional theory, numerical calculation based on a free-electron model as well as an analytical theory. Physical properties like electronic specific heat, spin susceptibility etc. depend on the electronic density of states (DOS) at the Fermi level. Therefore, these properties are expected to change with the layer thickness of ultrathin films near the two-dimensional limit, which is one atomic layer. This thickness dependence of DOS at Fermi energy provides tunability of materials properties in the 2-D to 3-D crossover regime. Consequently, materials properties can be tailored by appropriate electronic-structure engineering.

**Acknowledgement:** We acknowledge the financial support from IBIQuS project of the Department of Atomic energy.

**Figure Captions:**

FIG. 1: (a) A constant-current STM image showing Ag islands on a Si(111) surface. (b) Height profile along the line marked in (a).

FIG. 2: (a) A constant-current STM image showing Ag islands on a Si(111) surface ($V_g$ =0.5 V, $I_t$ = 0.5 nA). The island grows on top of a two-dimensional wetting layer. (b) High-resolution room temperature STM image of the groove, created by the tip-pulse voltage from the STM tip. STS measurements were made exactly on the marked positions. (c) Schematic cross-sectional view of the groove on top of the Ag island on Si(111) surface. At the marked positions in the slope of the groove STS measurements were performed by accessing different heights of the island. (d) Height profile across the groove along the line in (b) and its numerical derivative. The marked positions are the heights of the different layer thicknesses.

FIG. 3: (a) *I-V* curves obtained by STS measurements on the slope of the groove; the curves 1 to 6 correspond to different thicknesses. (b) The corresponding normalized tunneling conductances, (*dI/dV*)/(*I/V*).

FIG. 4: DOS at Fermi level: experimental, analytical and DFT results as a function of film thickness. The DOS at Fermi level is normalized to its bulk value.

FIG. 5: (a) Theoretical model for DFT calculation representing the sample ground state geometry. Top view and side view are shown in the upper and lower panel respectively. Four bilayers of silicon represent the substrate and 3 atomic layers Ag is grown on top of silicon. The other side is passivated by H atoms. (b) Total density of states of the model structure of Ag thin films of various thicknesses calculated from the DFT formalism. (c) Magnified view of (b) within the energy range ±1 eV.

**Figures:**

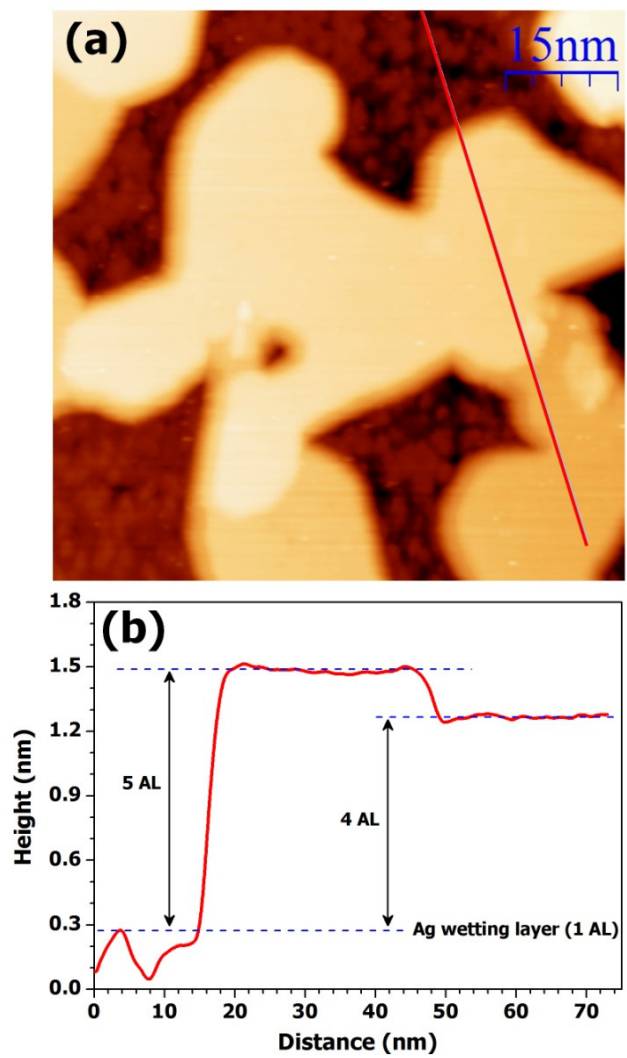

**FIG. 1**

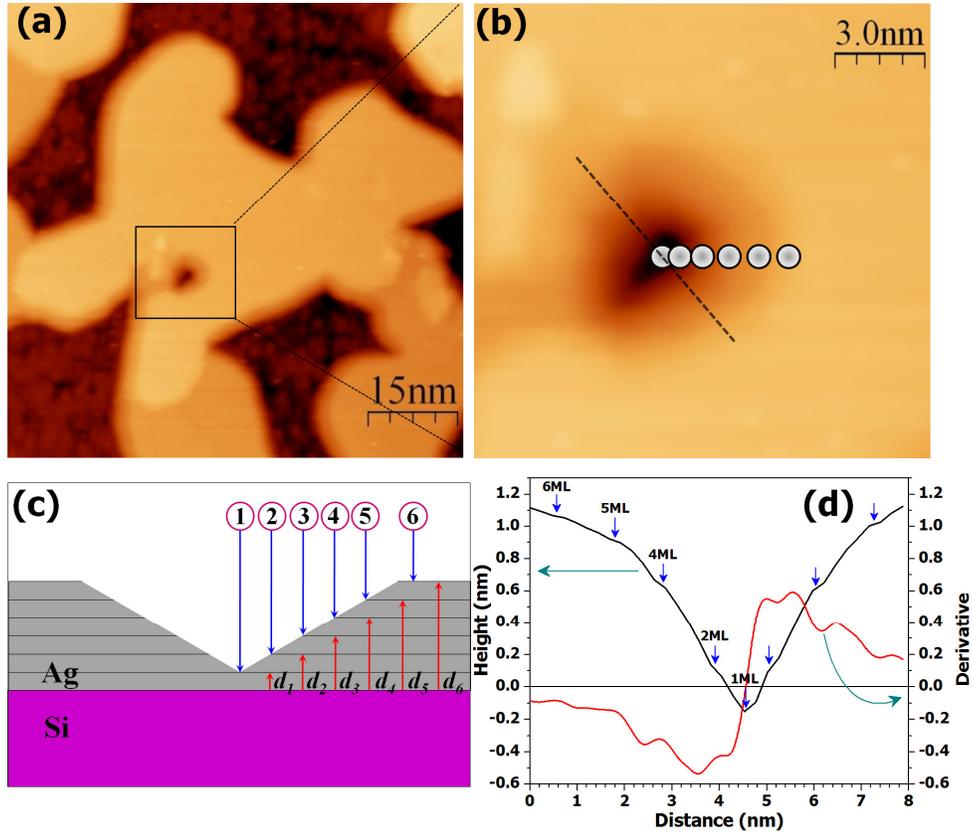

FIG. 2

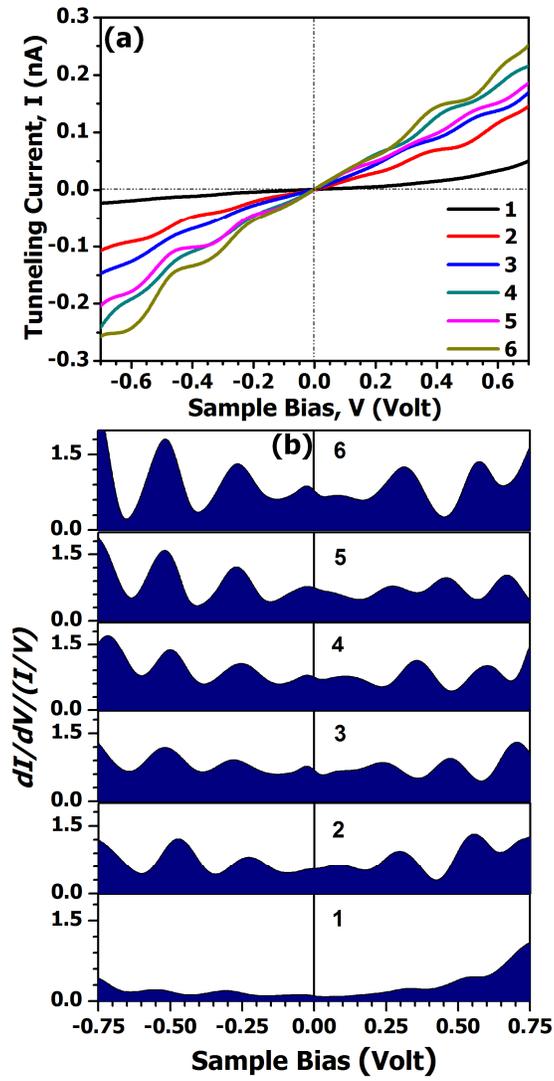

**FIG. 3**

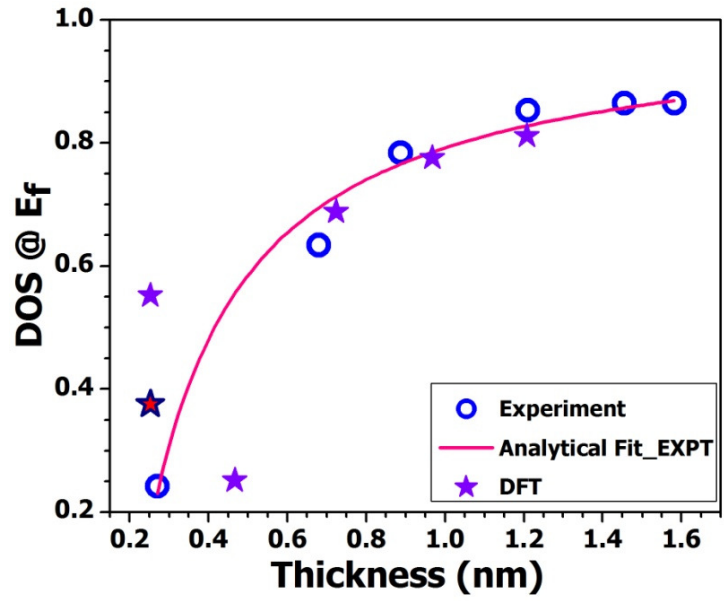

FIG. 4

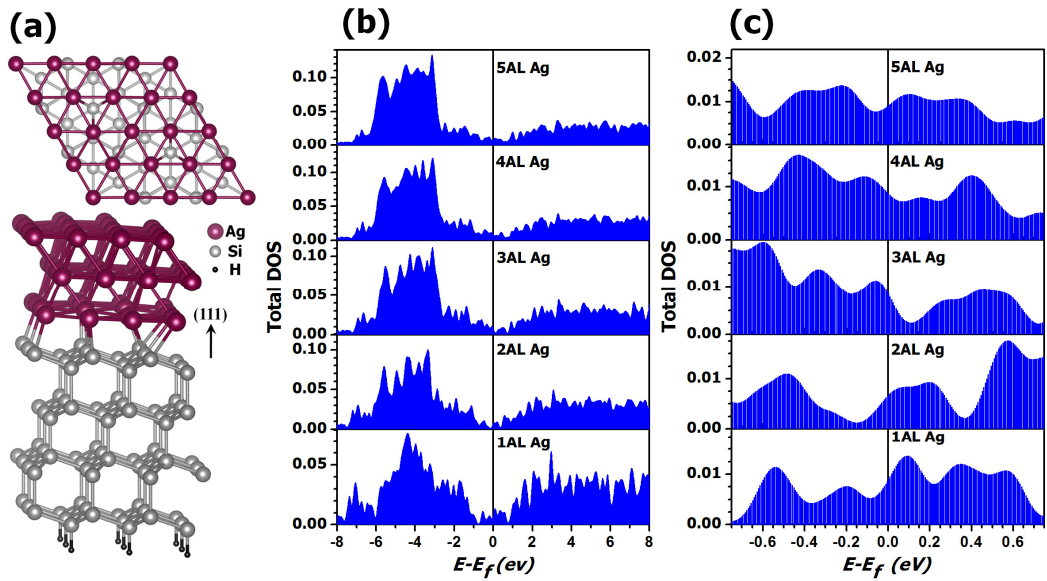

FIG. 5

**Supplementary Material:**

# Evolution of Fermi Level State Density in Ultrathin Films Near the Two–Dimensional Limit: Experiment and Theory

R. Batabyal, A. H. M. Abdul Wasey, J. C. Mahato, Debolina Das, G. P. Das and B. N. Dev*

Various materials properties depend on the electronic density of states (DOS) at Fermi energy, $\rho(E_F)$. For example, the electronic heat capacity ($C_{el}$) of a solid can be written as $C_{el} = \frac{\pi^2}{3}\rho(E_F)k_B^2 T$, where $k_B$ is the Boltzmann constant and $T$ is the absolute temperature [1]. Similarly, spin susceptibility, $\chi = \mu_B^2 \rho(E_F)$ (where $\mu_B$ is the Bohr magnetron) [1], Thomas-Fermi screening length, $l_{TF} = \left[e^2 \rho(E_F)/\varepsilon_0\right]^{-1/2}$ (where $\varepsilon_0$ is the dielectric constant and $e$ is the electronic charge) [2], etc. depend on DOS at Fermi energy.

**A. Analytical form of DOS at Fermi energy**

For ultrathin films, $\rho(E_F)$ would depend on film thickness [1, 3 - 4]. From an analytical model $\rho(E_F)$ can be written as [4]

$$\rho(E_F) = \rho^0\left(E_F^0\right)\left(1 - \frac{\gamma^0}{d_z}\right) \qquad (1)$$

where $\gamma^0 = \frac{\pi}{4k_F^0} - \frac{1}{2k_F^0}\left\{(2+\lambda^0)\sin^{-1}\left(\frac{1}{\sqrt{\lambda^0}}\right) - (\lambda^0 - 1)^{1/2}\right\}$,

$\rho^0(E^0_F)$ is the bulk value of DOS at Fermi energy, $d_z$ is the thickness of the infinitely large metal slab, $k_F^0$ is the Fermi wave vector, $k_{TOP} = (\sqrt{2mV_z}/\hbar)$, $\lambda^0 = (k_{TOP}/k^0_F)^2$, $V_z$ is the potential depth, $\hbar$ is Planck's constant and $m$ is the electron mass.

**B. DFT calculations for DOS at Fermi energy**

The DFT based Vienna *ab Initio* Simulation Package (VASP) [5] has been used for calculation. Projector augmented wave (PAW) [6] potentials were employed for elemental constituents, viz. Ag and Si potentials. Generalized gradient approximation (GGA) functional of Perdew, Burke and Ernzerhof (PBE) has been used [7] for calculating the exchange correlation energy. Brillouin zone sampling has been done using Monkhorst-Pack [8] method using 2×2×2 k-mesh. For all the calculations, self-consistency has been achieved with a 0.1 meV convergence in total energy. Cut off energy of 400 eV for the plane wave basis has been used. To obtain the optimized ground state geometry, 'forces' have been converged to less than 0.01 eV/Å by conjugate gradient minimization [9-10]. Layer projected DOS for the Ag layers and the Si bilayers are shown in Fig. S1.

Fig. S2(a) shows side view of the optimized ground state geometry of 1-AL Ag layer on Si(111). The arrow shows the [111] crystallographic direction of the system. A close look at the Ag adlayer reveals that the Ag atoms are not in the same plane, which is perpendicular to the [111] direction. That indicates a slight undulation in the Ag adlayer. This undulation of Ag layer right at the Ag/Si interface, i.e. at the very first layer of Ag in contact with the Si substrate, has been found to change as the layer thickness grows by addition of Ag layers one by one. Fig. S2(b) shows the undulation of Ag layer ($\delta z$ in Å, pink curve) right at the Ag/Si interface as a function of Ag layer thickness (in AL). The undulation is calculated by taking the difference between the position of the upper most Ag atoms and the position of the lower most Ag atoms in the first Ag layer. For 1-AL Ag, the value of $\delta z$ is found to be maximum. Our DFT calculation also indicates that the layer DOS at Fermi energy has a strong correlation with the undulation of the adlayer. Fig. S2(b) shows the layer-projected DOS at Fermi energy (blue curve) of the Ag layer right at the Ag/Si interface as a function of Ag coverage. This implies that the more the undulation of Ag adlayer at the Ag/Si interface, the more is the value of DOS at Fermi energy. Therefore, the undulation of Ag adlayer at the Ag/Si interface may, at least partly, be attributed to the origin of the deviation of the DOS at Fermi energy for 1-AL Ag (first data point in DFT results in Fig. 4). In experiment, the first Ag layer is at the bottom of the island and so the undulation is expected to be negligible. Accordingly, the adjusted DOS value for 1-AL Ag is shown in Fig. 4 as a star with an inscribed circle. This point is closer to the experimental value.

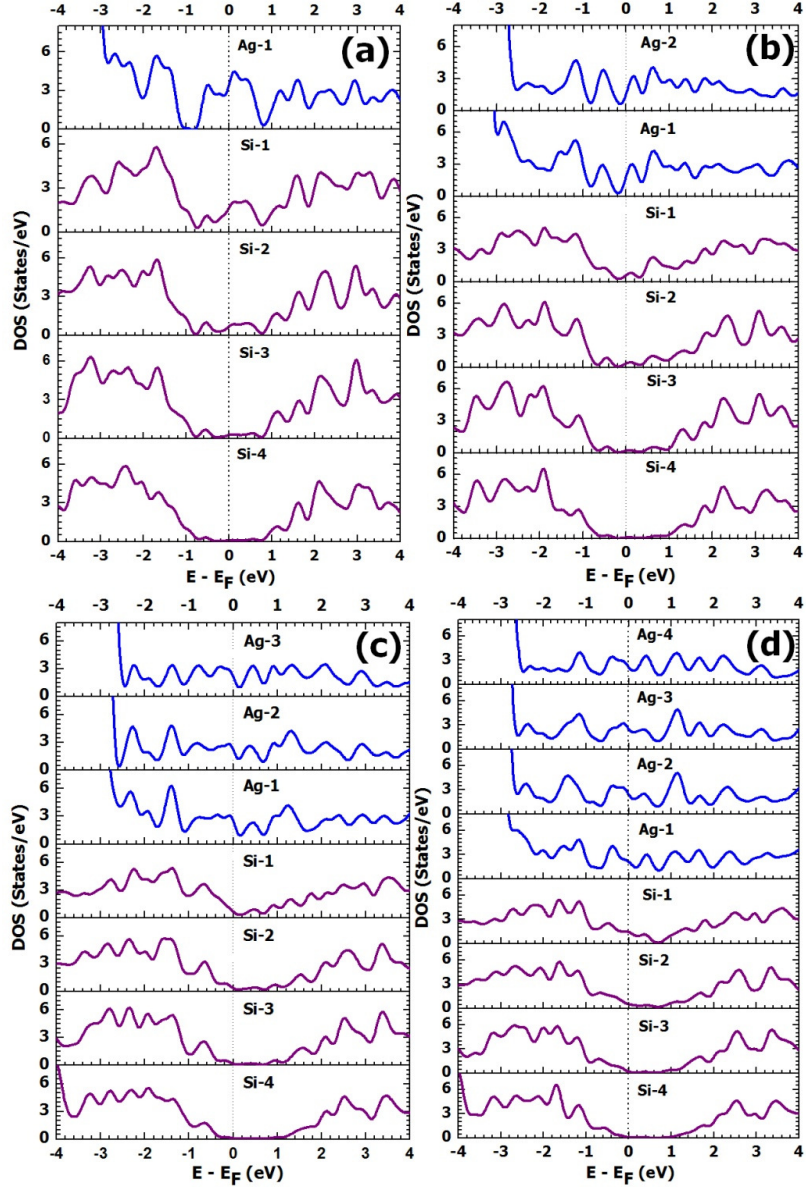

**FIG. S1:** (a) – (d) Calculated layer projected DOS of Ag layers starting from the one- to four atomic layers on four bilayers of Si(111) substrate. The other end of the substrate is passivated by H in each case.

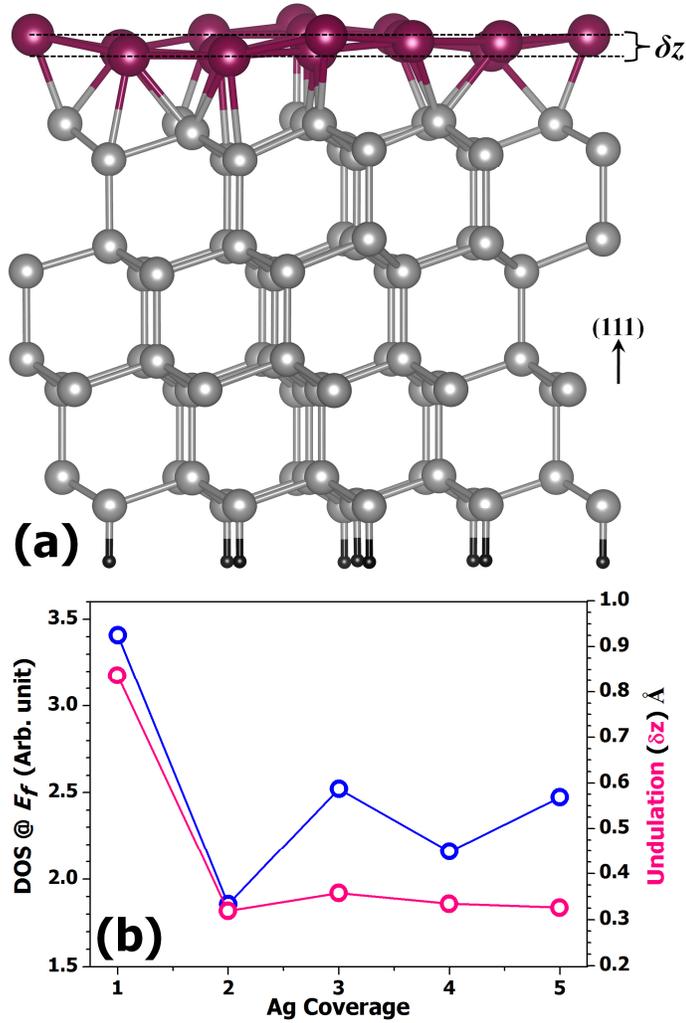

**FIG. S2:** DOS at Fermi energy calculated from layer projected DOS and undulation ($\delta z$) of the Ag atoms right at the interface layer as function of total Ag thickness in atomic layer. The undulation of the interface Ag atoms apparently correlates with the fluctuation in DOS at Fermi energy for the first layer.